\documentclass{iau_FM}
\usepackage{graphicx}

\title[The code \textsc{P\'egase}.3 for RG with JWST] 
{The code \textsc{P\'egase}.3 for distant radiogalaxies with the future JWST
} 

\author[Brigitte Rocca-Volmerange \& Michel Fioc]   
{Brigitte Rocca-Volmerange$^1$
\and Michel Fioc $^1$
 }
\affiliation{$^1$ Institut d'Astrophysique de Paris, CNRS  and Sorbonne Universit\'e, \\ 98bis Bd Arago,
F-75014, Paris France\\ email: {\tt rocca@iap.fr},  {\tt fioc@iap.fr}  \\[\affilskip]
 }

\pubyear{2018}
\jname{Astronomy in Focus, Volume 1} 
\editors{Volker Beckmann \& Claudio Ricci, eds.}
\begin{document}
\maketitle

\begin{abstract}
The physical link of the star formation-AGN activities is analyzed from multiwavelength energy distributions of distant radiogalaxies (RG) with the help of two models: the new evolutionary code \textsc{P\'egase}.3 with dust predictions and the Siebenmorgen's AGN model, with the aim to to disentangle the farIR dust emissions from respectively interstellar medium and torus of RGs. Best-fits of the HST-Spitzer-Herschel (UV-to-IR) observations of the 3CR RGs with libraries of hybrid SED templates identify three components (AGN, old galaxy and young starburst)  tracing the relation of starburst-AGN luminosities. To confirm this relation at higher resolutions, the JWST/NIRCam, MIRI and NIRspec instruments are needed, SED libraries and evolving colors of galaxy hosts adapted to the JWST instruments are in preparation with \textsc{P\'egase}.3.  
\keywords{galaxies:evolution, infrared: galaxies, (ISM:) dust, emission}
\end{abstract}

\section{The 3 components of RG: AGN, starburst and old galaxy}
\begin{figure}
  \noindent
  \begin{minipage}{0.6\linewidth}
    \includegraphics[width=\linewidth]{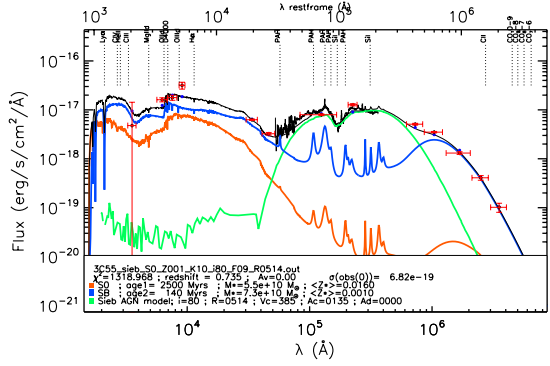} 
  \end{minipage}\hfill
  \begin{minipage}{0.4\linewidth}
    \caption{Best-fits of the HST-Spitzer-Herschel observations of the radio galaxy 3C~368 show three components: the old (orange) population of early-type galaxy, a recent starburst (blue) and the AGN torus (green).  The library of hybrid templates is built with \textsc{P\'egase}.3 and the clumpy Siebenmorgen's AGN model. The disentangling is consistent with the Spitzer/IRS spectra within the error bars.}
    \label{fig1Rocca}
  \end{minipage}
\end{figure}

Multiwavelength hybrid SEDs for distant radiogalaxies are built by associating  the evolution of a massive host galaxy and of a short intense starburst from the code \textsc{P\'egase}.3, with the constant far-IR emission of a clumpy AGN torus model (\cite{Drouart2016}). \textsc{P\'egase}.3 (Fioc \& Rocca-Volmerange, submitted to Astronomy  \& Astrophysics Journal,  see also www2.iap.fr/pegase) predicts from 0 to 20 Gyrs, the stellar and nebular emissions, corrected for metallicity dependent dust attenuation, updated from  \cite{Pegase2},  extended to the coherent dust emission  of the same source respecting the energy balance by radiative transfer MonteCarlo simulations. The best-fit procedures of these multiple synthetic libraries on UV-to-far-IR observations  of 3CR galaxies (Fig.\,\ref{fig1Rocca}) identify the three components (AGN, old and young stellar populations). 
\section{The 3CR catalog: a star formation-AGN link?}
\begin{figure}
\vspace{-5mm}
\begin{minipage}{0.55\linewidth}
  \includegraphics[width=\linewidth]{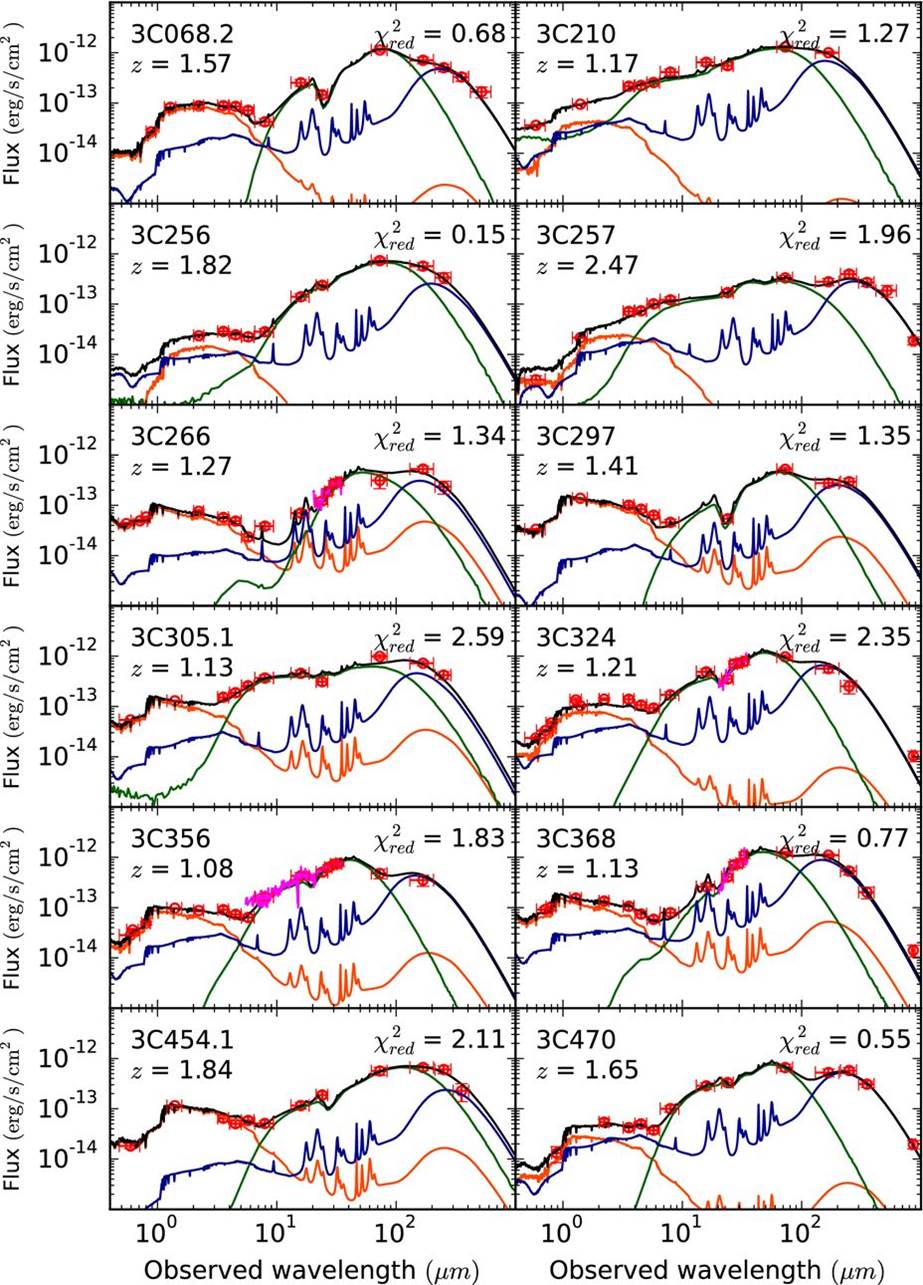} 
\end{minipage}\hfill
\begin{minipage}{0.35\linewidth}
  \caption{ 3CR radiogalaxies show the same distribution of the three components (AGN, old and recent) at various redshifts.}
  \label{fig2Rocca}
\end{minipage}
\end{figure}

\begin{figure}
  \noindent
  \begin{minipage}{0.6\linewidth}
 \includegraphics[width=0.60\linewidth]{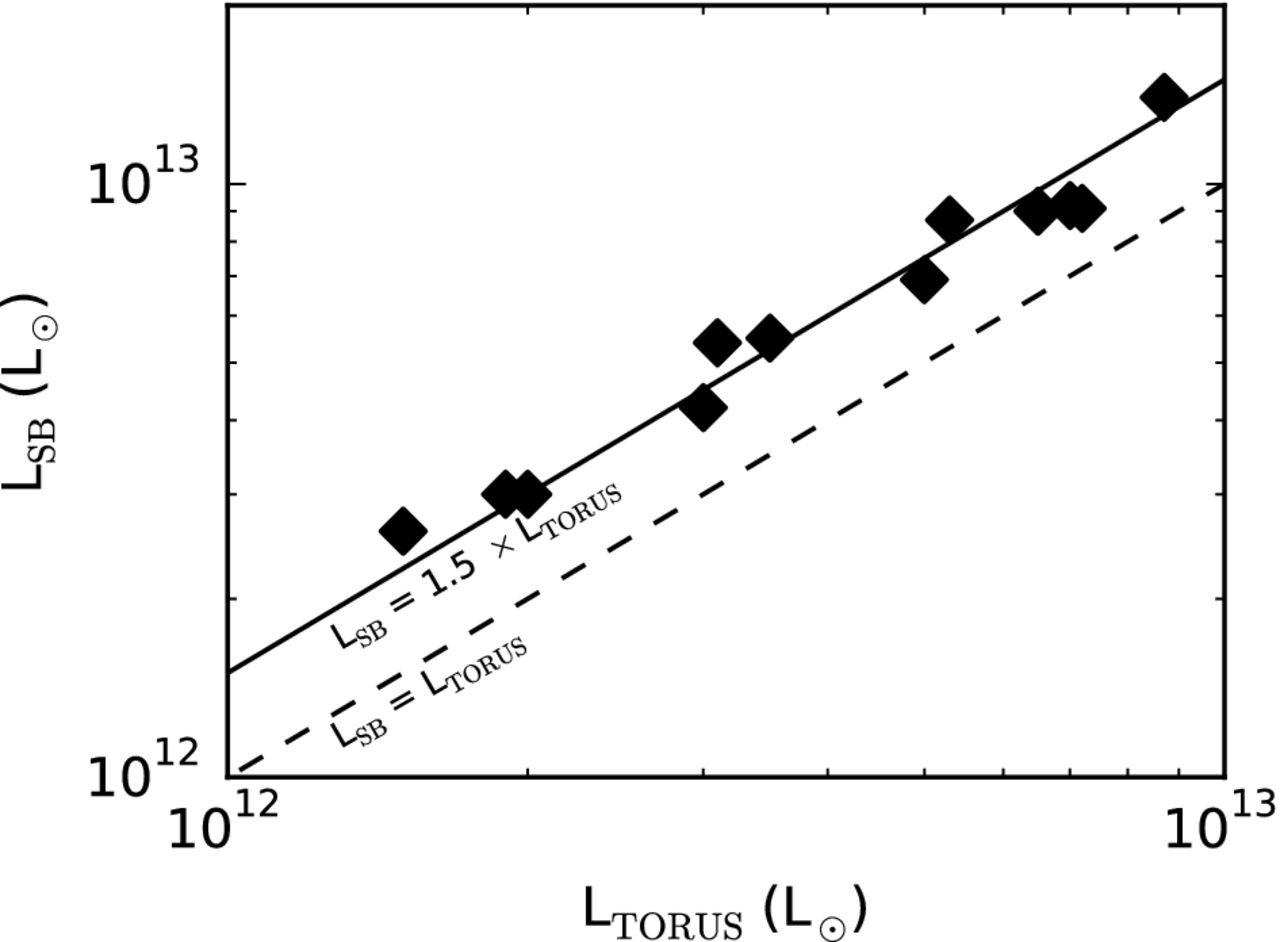}
\end{minipage}\hfill
\begin{minipage}{0.6\linewidth}
  \caption{ The luminosity relation of $L_{\mathrm{starburst}}$ with $L_{\mathrm{AGN}}$ for the analyzed 3CR galaxies.}
  \label{fig3Rocca}
\end{minipage}
\end{figure}

\cite{Podigachoski2016} applied the hybrid method to galaxies of the 3CR catalog at various z. The three components are found in all cases (Fig.\,\ref{fig2Rocca}), even at higher redshifts.
 From the 3CR galaxies, an interesting relation is observed between the luminosities $L_{\mathrm{starburst}}$ and  $L_{\mathrm{AGN}}$, respectively from the dusty starburst and the AGN torus (Fig.\,\ref{fig3Rocca}),  suggesting a relation between star formation and AGN activities hinted by the different thermal peaks in the far-IR. In the mid-IR, these components deserve to be analyzed with better spatial and spectral resolutions of the {\it JWST/NIRcam, MIRI, NIRspec} instruments. We plan to build \textsc{P\'egase}.3 template libraries and  corresponding synthetic colors to help the community to analyze future JWST data. 

\end{document}